\documentclass[aps, prl, twocolumn, preprintnumbers, superscriptaddress, showpacs, amssymb]{revtex4-1}

\usepackage{graphicx}
\usepackage{dcolumn}
\usepackage{bm}

\begin{document}

\preprint{Ver. 3p1}¡¡

\title{Magnetotransport study of the pressure-induced antiferromagnetic phase in FeSe}


\author{Taichi Terashima}
\author{Naoki Kikugawa}
\affiliation{National Institute for Materials Science, Tsukuba, Ibaraki 305-0003, Japan}
\author{Shigeru Kasahara}
\author{Tatsuya Watashige}
\author{Yuji Matsuda}
\affiliation{Department of Physics, Kyoto University, Kyoto 606-8502, Japan}
\author{Takasada Shibauchi}
\affiliation{Department of Advanced Materials Science, University of Tokyo, Chiba 277-8561, Japan}
\author{Shinya Uji}
\affiliation{National Institute for Materials Science, Tsukuba, Ibaraki 305-0003, Japan}


\date{\today}
\begin{abstract}
The resistivity $\rho$ and Hall resistivity $\rho_H$ are measured on FeSe at pressures up to $P$ = 28.3 kbar in magnetic fields up to $B$ = 14.5 T.
The $\rho(B)$ and $\rho_H(B)$ curves are analyzed with multicarrier models to estimate the carrier density and mobility as a function of $P$ and temperature ($ T\leqslant$ 110 K).
It is shown that the pressure-induced antiferromagnetic transition is accompanied by an abrupt reduction of the carrier density and scattering.
This indicates that the electronic structure is reconstructed significantly by the antiferromagnetic order.
\end{abstract}

\pacs{74.70.Xa, 72.15.Gd, 74.25.Dw, 74.25.Jb}

\maketitle



\newcommand{\ud}{\mathrm{d}}
\def\degree{\kern-.2em\r{}\kern-.3em}

Since the discovery of superconductivity (SC) at $T_c$ = 26 K in LaFeAs(O$_{1-x}$F$_x$) by Kamihara \textit{et al.} \cite{Kamihara08JACS}, the iron-based high-$T_c$ materials have intensively been studied.
Yet, the paring mechanism is still highly controversial: some propose spin fluctuations as the glue \cite{Mazin08PRL, Kuroki08PRL}, while others orbital ones \cite{Kontani10PRL, Yanagi10PRB}.
Intimately related to this issue is the origin of the nematic transition \cite{Fernandes14NatPhys}.
Typical iron-pnictide parent compounds such as LaFeAsO or BaFe$_2$As$_2$ \cite{Rotter08PRL, Sasmal08PRL} exhibit a tetragonal-to-orthorhombic structural transition at $T_s$, slightly above or at the same temperature as a stripe-type antiferromagnetic (AFM) order at $T_N$.
Electronic properties below $T_s$ exhibit in-plane anisotropy that is much stronger than expected from the slight orthorhombic distortion \cite{Chu10Science, Dusza11EPL, Nakajima11PNAS, Yi11PNAS}.
It is therefore believed that the structural transition is driven by electronic (i.e., spin or orbital) degrees of freedom and thus it is called a nematic transition.
$T_s$ and $T_N$ are reduced simultaneously by pressure or chemical substitution, resulting in a phase diagram where the AFM phase is enclosed by the nematic one \cite{Chu10Science}.
This is consistent with scenarios of spin-driven nematicity \cite{Fernandes14NatPhys, Yamase15NJP}.
An SC dome appears around points where $T_s$ and $T_N$ reach zero.

\begin{figure}
\includegraphics[width=7cm]{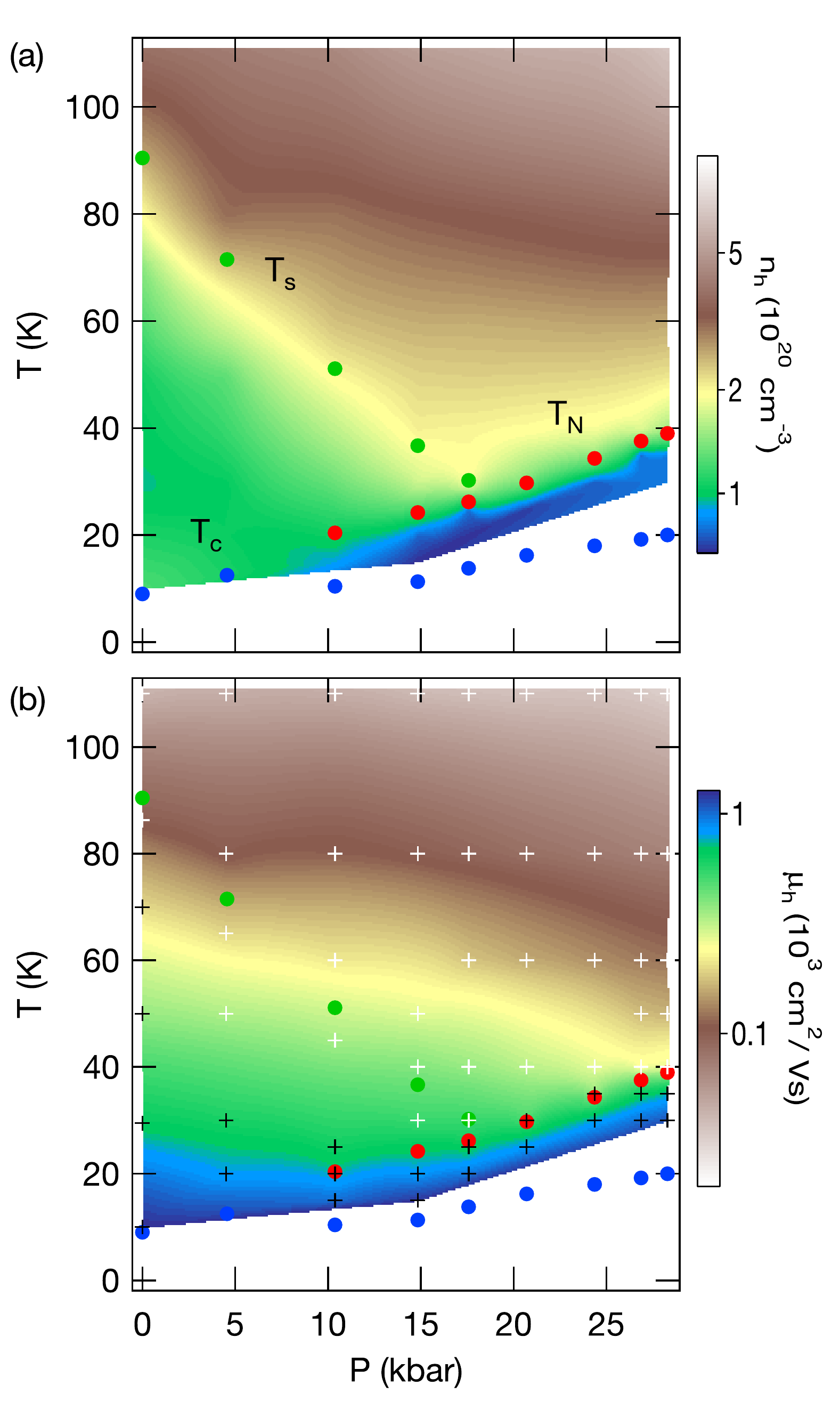}
\caption{\label{color_plot}(Color online).  (a) Carrier density $n_h = n_{e1} + n_{e2}$ and (b) hole mobility $\mu_h$ as a function of pressure and temperature.  Structural ($T_s$, green online), AFM ($T_N$, $T_u$ in \cite{Terashima15JPSJ}, red online), and SC ($T_c$, blue online) transition temperatures for the present sample are also shown.  $T_c$ is defined as the temperature where the resistance becomes zero.  White (Black) crosses in (b) indicate points where the compensated two(three)-carrier fit was employed.}   
\end{figure}

FeSe ($T_c \approx$ 8 K \cite{Hsu08PNAS}) initially attracted attention because of a remarkable enhancement of $T_c$ by pressure \cite{Mizuguchi08APL, Medvedev09Nmat, Margadonna09PRB, Garbarino09EPL}.
A report of $T_c > 50$ K in single-layer films \cite{Wang12CPL} has also aroused considerable interest.
At the same time, FeSe may be crucial in determining the paring glue and the origin of the nematicity in the iron-based superconductors.
It undergoes a structural transition at $T_s$ $\sim$ 90 K but does not order magnetically at ambient pressure \cite{McQueen09PRL}.
A large splitting of the $d_{xz}$ and $d_{yz}$ bands below $\sim T_s$ found by angle-resolved photoemission spectroscopy indicates that the transition at $T_s$ is a nematic one accompanied by orbital polarization \cite{Tan13NatMat, Shimojima14PRB, Nakayama14PRL}.
An AFM transition can be induced by pressure \cite{Bendele10PRL, Bendele12PRB, Terashima15JPSJ}.
However, the phase diagram (see Fig.~\ref{color_plot} and \cite{Terashima15JPSJ, Kaluarachchi16PRB, Sun15condmat}) is distinct from the typical one described above for the iron-pnictide compounds.
This casts some doubts on the spin-nematic scenarios.
It is however to be noted that low-energy AFM spin fluctuations have been observed below $T_s$ in NMR and inelastic neutron scattering measurements \cite{Imai09PRL, Baek14nmat, Bohmer15PRL, Rahn15PRB, Wang15NatMater, Shamoto15condmat}.
In addition, very recent reports suggest that at high pressures ($P\gtrsim$ 18 kbar, Fig.~\ref{color_plot}) where no separate structural transition at $T_s$ is seen the AFM transition at $T_N$ is accompanied by an orthorhombic distortion \cite{Kothapalli16condmat, Wang16condmat}.

It is important to know how the two transitions at $T_s$ and $T_N$ alter the electronic structure.
So far Shubnikov-de Haas (SdH) measurements have been performed at ambient pressure \cite{Terashima14PRB, Audouard15EPL, Watson15PRB, Watson15PRL} and under high pressure \cite{Terashima16PRB}.
The Fermi surface (FS) at ambient pressure is anomalous, deviating significantly from that predicted by band-structure calculations \cite{Terashima14PRB}.
The carrier density is one order-of-magnitude smaller than calculated.
In the high-pressure measurements up to 16.1 kbar \cite{Terashima16PRB}, a drastic change in SdH oscillations has been found to occur at the onset of the AFM order at $P \sim$8 kbar: high SdH frequencies, corresponding to large FS cross sections, suddenly disappear, and only a low frequency remains.
This suggests that parts of the FS is gapped by a reconstruction due to the AFM order.
In this work, we measure the electrical resistivity and Hall effect in FeSe under high pressure.
The magnetotransport data confirm that the carrier density is reduced in the AFM phase, and also show that the carrier scattering is reduced.

\begin{figure}
\includegraphics[width=7cm]{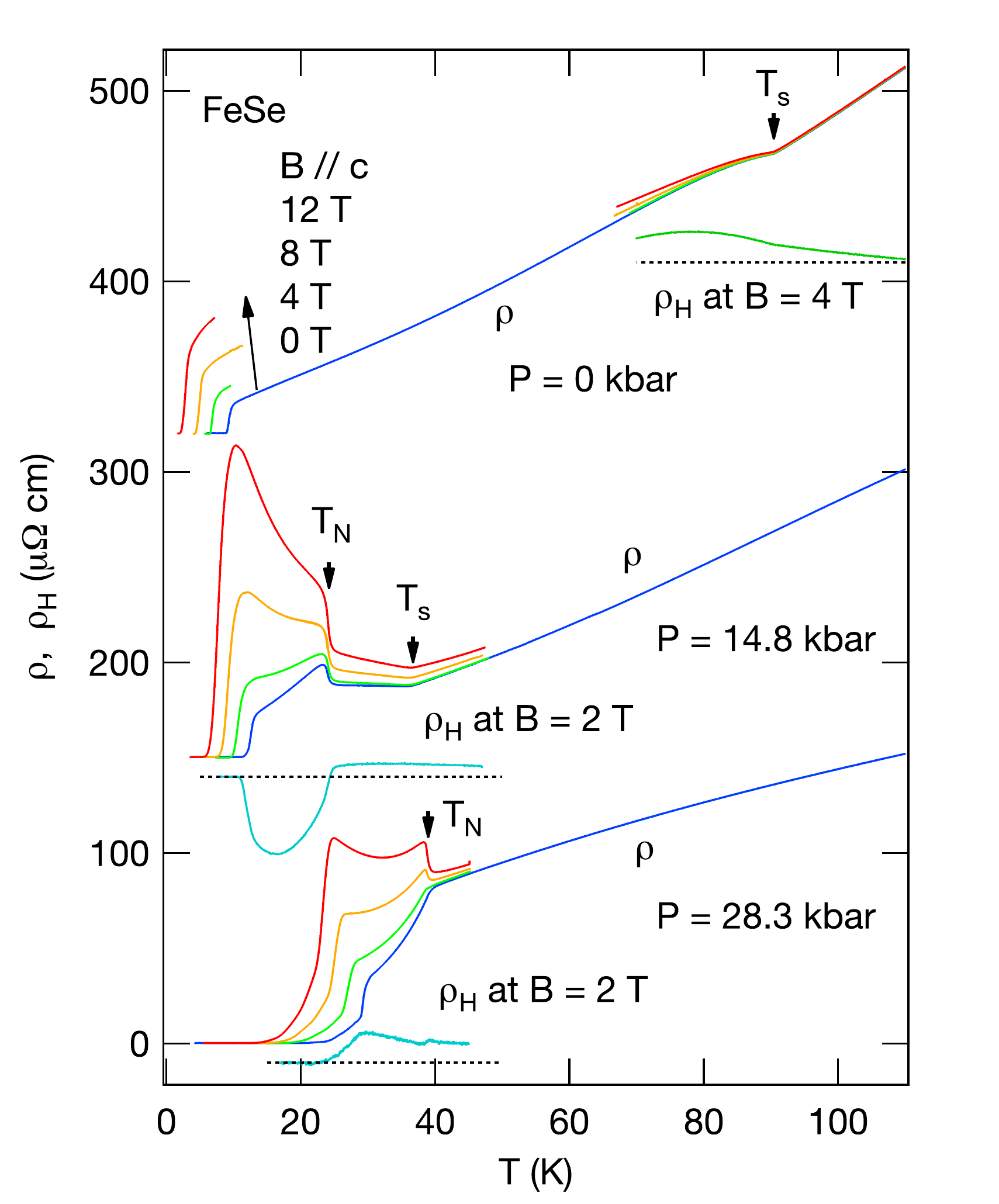}
\caption{\label{RvsT}(Color online).  Temperature dependence of the resistivity $\rho$ in FeSe for selected pressures.  The resistivity in applied magnetic fields and the Hall resistivity $\rho_H$ at a constant field are also shown for temperature regions near phase transitions.  Curves are vertically shifted for clarity, and zero for each of the $\rho_H$ curves is indicated by a broken line.}   
\end{figure}

The single crystal was prepared in Kyoto by a chemical vapor transport method \cite{Bohmer13PRB}, which is known to produce highly stoichiometric crystals with good crystallinity \cite{Bohmer13PRB, Kasahara14PNAS, Watashige15PRX}.
The dimensions, $T_c$ (zero resistance), and the resistivity ratio between $T$ = 11 K and room temperature are 1.8 $\times$ 1.0 $\times$ 0.05 mm$^3$, 9.0 K, and 29, respectively \cite{SM}.
The resistivity $\rho$ and Hall resistivity $\rho_H$ were measured down to $\sim$2 K in magnetic fields up to $B$ = 14.5 T applied parallel to the $c$ axis.
Six electrical contacts were spot-welded in the standard geometry, and a low-frequency ac current ($f$ = 11.3 Hz, $I$ $\leqslant$ 3 mA) was applied in the $ab$ plane.
The longitudinal (transverse) voltage was (anti)symmetrized with respect to the field to obtain $\rho$ ($\rho_H$), except for $\rho(T)$ curves in Fig.~\ref{RvsT}, for which no symmetrization was applied.
A piston-cylinder type pressure cells made of NiCrAl alloy (C\&T Factory, Tokyo) \cite{Uwatoko02JPCM} was used to generate pressures up to $P$ = 28.3 kbar.
The pressure transmitting medium was Daphne 7474 (Idemitsu Kosan, Tokyo), which remains liquid up to 37 kbar at room temperature and assures highly hydrostatic pressure generation in the present pressure range \cite{Murata08RSI, Klotz09JPhysD, Tateiwa09RSI, Terashima16PRB}. 
The pressure was determined from the resistance variation of calibrated manganin wires.
The applied pressures and corresponding phase transition temperatures can be seen in Fig.~\ref{color_plot}.
The upper critical field data are given in \cite{SM}.

The present resistivity and ambient-pressure Hall data agree well with previous reports \cite{Terashima15JPSJ, Kasahara14PNAS, Huynh14PRB, Watson15PRL, KnonerPRB15, RosslerPRB15, Kaluarachchi16PRB, Sun16PRB, Sun15condmat}.
To our knowledge, no high-pressure Hall data has been reported so far.
The deduced phase diagram (Fig.~\ref{color_plot}) is in excellent agreement with those reported in \cite{Terashima15JPSJ, Kaluarachchi16PRB, Terashima16PRB}, guaranteeing that the present data reveal intrinsic behavior of high-quality FeSe crystals.
Note that $T_s \approx 80$ K at ambient pressure reported in \cite{Kang16SST} is significantly lower than other reports \cite{Bohmer13PRB, Bohmer15PRL, Terashima15JPSJ, Kasahara14PNAS, Huynh14PRB, KnonerPRB15, RosslerPRB15, Kaluarachchi16PRB, Sun16PRB, Sun15condmat} and hence that the phase diagram in \cite{Kang16SST} do not agree with ours.

Figure~\ref{RvsT} shows the temperature dependences of $\rho$ and $\rho_H$ for three selected pressures.
At $P$ = 0 kbar, $\rho(T)$ shows a kink at $T_s$ = 90.5 K (defined by a positive peak of d$^2\rho$/d$T^2$), which stays the same within experimental accuracy when the field is applied up to 12 T.
The magnetoresistance (MR) is enhanced gradually with decreasing temperature below $T_s$.
$\rho_H(T)$ exhibits only a slight change in the slope at $T_s$ and then reaches a broad maximum at $\sim$80 K.
The orthorhombic distortion without a stripe magnetic order only slightly distorts the Brillouin zone: it does not cause a band folding.
The accompanying change in the electronic structure is therefore expected to be small, which is consistent with the weak structure in $\rho_H(T)$ at $T_s$.

At $P$ = 14.8 kbar, $\rho(T)$ shows a kink at $T_s$ = 36.7 K and then jumps at $T_N$ = 24.2 K (defined by a negative peak of d$\rho$/d$T$).
Neither $T_s$ nor $T_N$ shifts with applied fields.
The insensitivity of $T_N$ indicates that the suppression of the magnetic susceptibility below $T_N$ is negligible, which might indicate that the AFM-ordered spins lie in the $ab$ plane.
The resistivity jump at $T_N$ becomes more prominent with increasing field, and the MR is considerably enhanced below $T_N$. 
$\rho_H(T)$ shows hardly any anomaly at $T_s$.
It however shows a clear bend at $T_N$ and shifts toward the negative side quickly with decreasing temperature, indicating that the electronic structure is reconstructed significantly at $T_N$.
The enhanced MR below $T_N$ suggests an enhanced carrier mobility.
Nevertheless, the zero-field $\rho$ increases just below $T_N$.
This most likely indicates that the suppression of the carrier density below $T_N$, which is suggested by $\rho_H$, dominates over the enhanced mobility, as we confirm below with multicarrier analyses.

At $P$ = 28.3 kbar, where no separate transition at $T_s$ is found, $\rho(T)$ at $B$ = 0 T exhibits a drop rather than a jump at $T_N$ = 39.0 K (determined from a negative peak of d$^2\rho$/d$T^2$ for this particular case).
As the field is applied, a jump appears at $T_N$ gradually, but the temperature $T_N$ stays the same.
The MR is considerably enhanced below $T_N$. 
$\rho_H(T)$ shows a kink at $T_N$.
Although this change at $T_N$ is not so prominent as that at $P$ = 14.8 kbar, it is largely due to the choice of a field of $B$ = 2 T.
As shown below, $\rho_H$ at higher fields is markedly enhanced below $T_N$. 

\begin{figure}
\includegraphics[width=8cm]{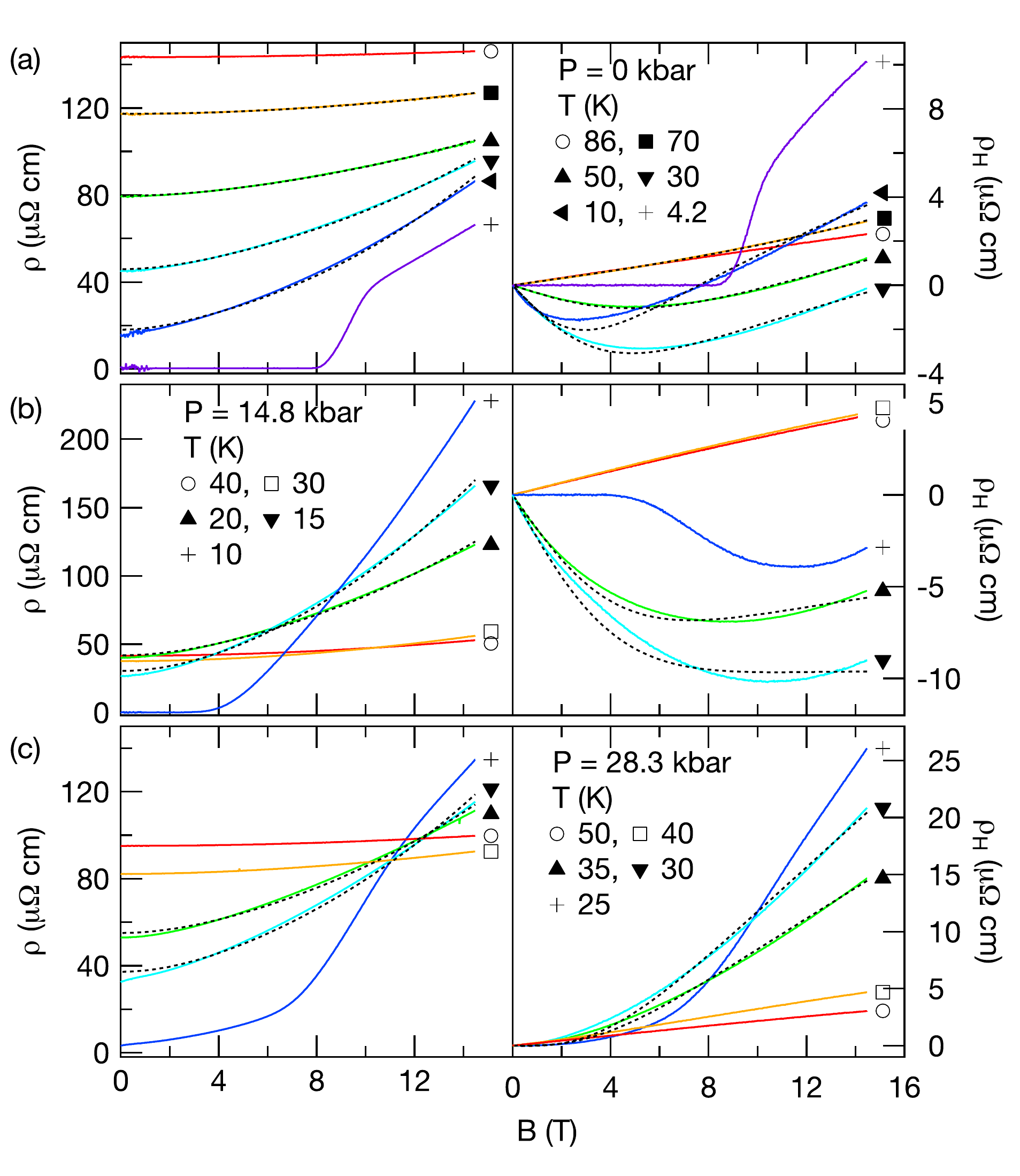}
\caption{\label{RvsB}(Color online).  Magnetic-field dependence of the resistivity $\rho$ (left) and Hall resistivity $\rho_H$ (right) in FeSe for selected pressures and temperatures.  Broken lines show compensated three-carrier fits (see text).  Note $T_s$ = 90.5 K at $P$ = 0 kbar (a), $T_s$ = 36.7 K, $T_N$ = 24.2 K at $P$ = 14.8 kbar (b), and $T_N$ = 39.0 K at $P$ = 28.3 kbar (c).}   
\end{figure}

\begin{figure}
\includegraphics[width=7cm]{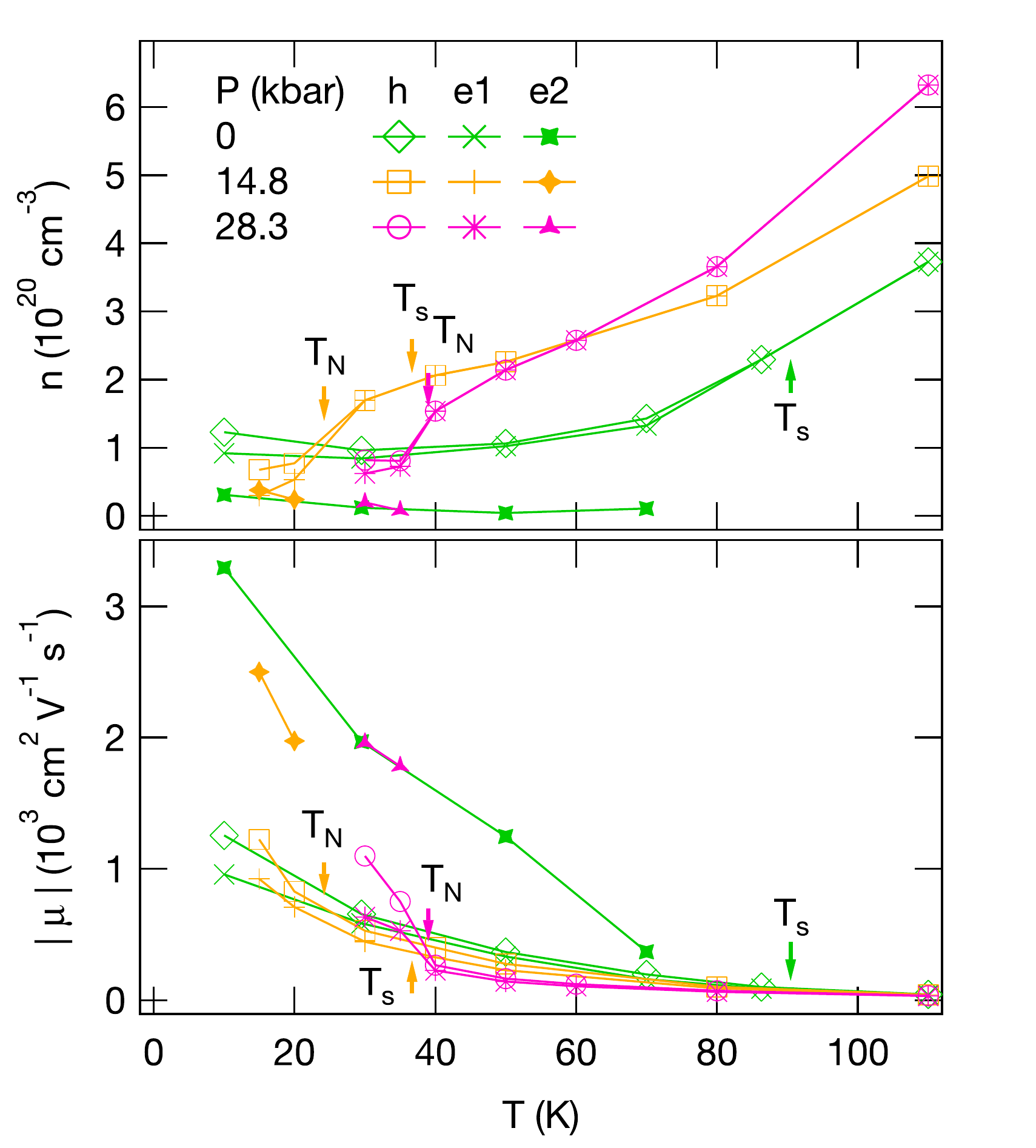}
\caption{\label{n_mu}(Color online).  Carrier densities and mobilities as a function of temperature for selected pressures.  A compensated two-carrier (hole h and electron e$_1$) and compensated three-carrier model (hole h and electrons e$_1$ and e$_2$) have been used to simultaneously fit resistivity and Hall resistivity data at high and low temperatures, respectively.}   
\end{figure}

Figure~\ref{RvsB} shows the magnetic-field dependences of $\rho$ and $\rho_H$ at selected temperatures for the above three pressures.

At $P$ = 0 kbar [Fig.~\ref{RvsB}(a)], $\rho_H(B)$ is linear at $T$ = 86 K and above, which can be explained with a compensated two-carrier model \cite{SM}.
$\rho_H(B)$ is nonlinear at lower temperatures.
Note that the onset of the nonlinearity does not coincide with $T_s$ (= 90.5 K).
The MR is enhanced not suddenly at $T_s$ but gradually with decreasing temperature.
In previous reports \cite{Huynh14PRB, Watson15PRL}, the nonlinear $\rho_H(B)$ has been ascribed to the existence of a third carrier with a high mobility and has been explained with a compensated three-carrier model \cite{Watson15PRL}.
The model consists of one hole (h) and two electron carriers (e1 and e2), and one of the two electron carriers (e2) has a larger mobility \cite{SM}. 
It can explain the present data satisfactorily: the broken lines in the figure are results of simultaneous fits of $\rho(B)$ and $\rho_H(B)$ to the model (the data at $T$ = 4.2 K was not analyzed because of SC).
The carrier densities and mobilities estimated from the compensated two- and three-carrier fits at high and low temperatures, respectively, are shown in Fig.~\ref{n_mu}.
The carrier density $n_h = n_{e1}+n_{e2}$ decreases gradually with decreasing temperature from above $T_s$ but increases below 30 K.
The carrier mobilities increase gradually with decreasing temperature.
The mobility of the second electron carrier $\mu_{e2}$ is about three times larger than the others.
These observations are consistent with previous reports \cite{Huynh14PRB, Watson15PRL}.

\begin{figure}
\includegraphics[width=7cm]{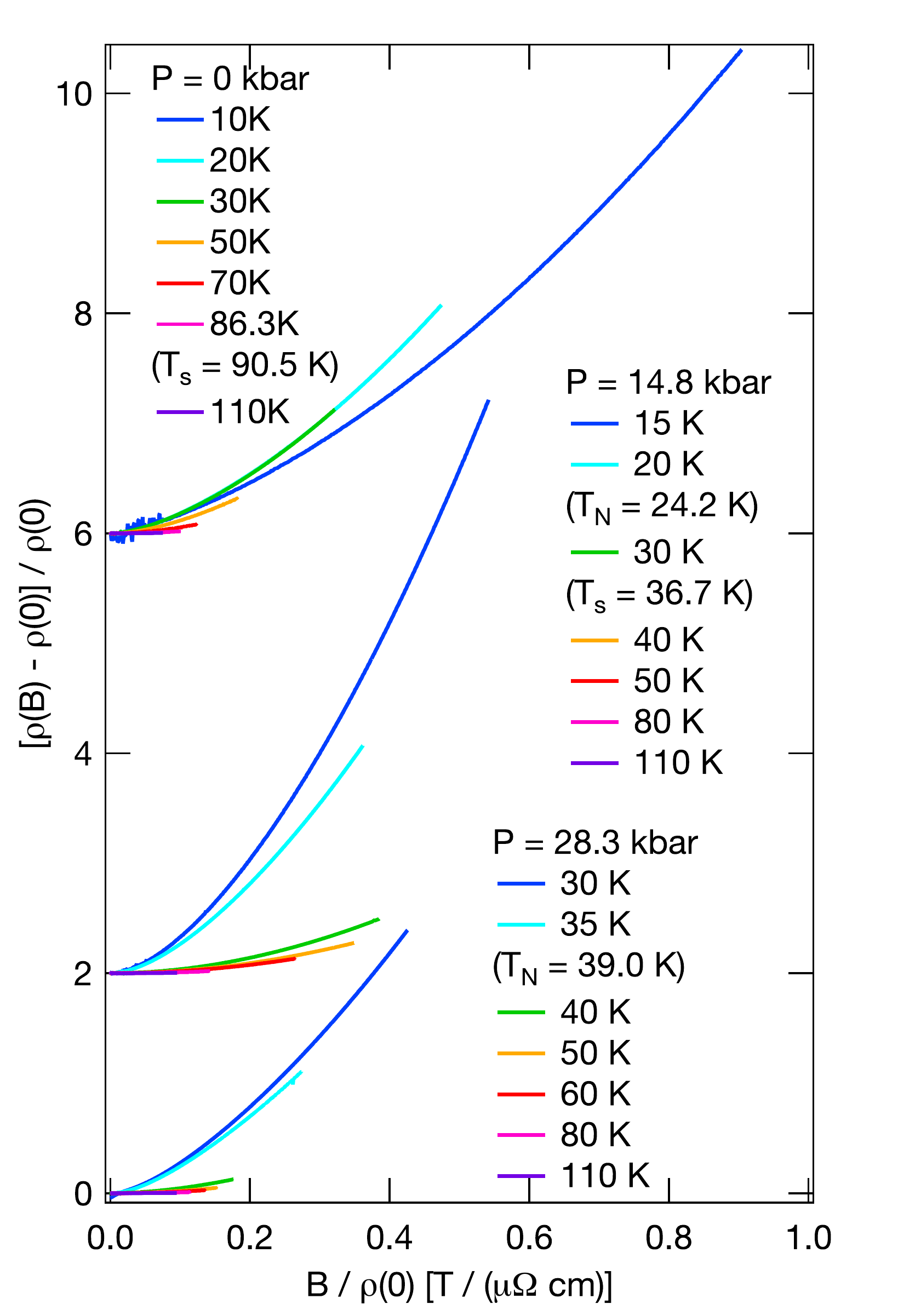}
\caption{\label{Kohler}(Color online).  Kohler plots for selected pressures.  Curves are offset for clarity.  The highest field is $B$ = 14.5 T for all the curves.}   
\end{figure}

At $P$ = 14.8 kbar, sudden changes occur in $\rho(B)$ and $\rho_H(B)$ across $T_N$ = 24.2 K, not $T_s$ = 36.7 K: compare the $T$ = 30 and 20 K curves in Fig.~\ref{RvsB}(b).
$\rho_H(B)$ remains linear down to $T$ = 30 K ($< T_s$) but is nonlinear at and below $T$ = 20 K ($< T_N$).
The MR is markedly enhanced below $T_N$ (see also Fig.~\ref{Kohler}).
These sudden changes are very different from the gradual ones observed at $P$ = 0 kbar.
The nonlinear $\rho_H(B)$ was previously observed in the AFM phase of the 122 iron-pnictide parent compounds such as BaFe$_2$As$_2$ and EuFe$_2$As$_2$ and was explained with the same compensated three-carrier model as above \cite{Ishida11PRB, Kurita13PRB}.
It can also explain the present data as shown by the broken lines in the figure (the data at $T$ = 10 K was not analyzed because of SC).
The carrier density $n_h$ shows an abrupt drop at $T_N$ and continues to decrease down to the lowest temperature of 15 K (Fig.~\ref{n_mu}).
The mobilities are enhanced below $T_N$.

At $P$ = 28.3 kbar [Fig.~\ref{RvsB}(c)], as the temperature is lowered below $T_N$ = 39.0 K, $\rho_H(B)$ becomes nonlinear, and the MR is strongly enhanced.
Those changes are as sudden as observed at $P$ = 14.8 kbar.
The data below $T_N$ can also be fitted to the compensated three-carrier model.
Figure~\ref{n_mu} shows that the carrier density drops abruptly at $T_N$ and that the mobilities are markedly enhanced below $T_N$.

Figure~\ref{Kohler} shows Kohler plots for the three pressures.
The resistivities at ($P$, $T$) = (0 kbar, 10 K) and (28.3 kbar, 30 K) are partially decreased by SC near $B$ = 0.
Thus the corresponding $\rho(0)$ values for the Kohler plots have been estimated by extrapolating the $\rho(T)$ curves from slightly higher temperatures and are 3 and 5\% larger than the measured resistivities at $B$ = 0, respectively.
Kohler's rule is not obeyed in any phase at any of the investigated pressures in the investigated temperature range ($T \leqslant$ 110 K, above which the MR is too small to verify the rule).
Previous ambient-pressure studies \cite{RosslerPRB15, Sun16PRB} have concluded that the rule is obeyed for a low temperature region $T \leqslant$ 30 K.
However, the present data at $P$ = 0 kbar indicates that the rule holds fortuitously only in a narrow temperature region $T \approx$ 20--30 K: the $T$ = 10 K curve clearly deviates from the 20 and 30 K curves (Fig.~\ref{Kohler}).

The violation of Kohler's rule indicates that the carrier scattering rate and its temperature evolution are significantly anisotropic on the FS.
In such a case, the simple multicarrier models used above do not hold quantitatively \cite{Kemper11PRB, Breitkreiz14PRB}: estimated carrier densities may deviate from actual ones. 
An intuitive explanation for this would be that parts of the FS where scattering is particularly strong, dubbed hot spots, contribute little to the electrical transport and hence they are missed, although a recent theory suggests a more involved explanation \cite{Breitkreiz14PRB}.
In fact, the actual carrier density at $P$ = 0 kbar as $T \to 0$ is estimated from SdH oscillations \cite{Terashima14PRB} to be $\sim3\times10^{20}$ cm$^{-3}$ \footnote{ARPES data at $T$ = 10 K \cite{Watson15PRB} give a similar estimate.}, approximately three times larger than the present estimate at $T$ = 10 K (Fig.~\ref{n_mu}), as pointed out in \cite{Watson15PRL}.
The present estimated density at $P$ = 0 kbar increases with decreasing temperature below 30 K.
This may be regarded as asymptotic behavior toward the true density because the scattering rate becomes less anisotropic as $T \to 0$.
In the pressure-induced AFM phase, the AFM spin fluctuations, the main source of the anisotropic scattering, are suppressed, and hence the carrier density is expected to be better estimated.
In high-pressure SdH measurements \cite{Terashima16PRB}, only a small frequency of $F \approx$ 0.1--0.2 kT has been observed in the AFM phase.
If the SdH oscillation can be ascribed to a two-dimensional FS cylinder, the carrier density will be 1--2$\times10^{20}$ cm$^{-3}$.
However, considering that the FS in the AFM phase of BaFe$_2$As$_2$ is three dimensional and closed \cite{Terashima11PRL}, it is more likely that the oscillation comes from a closed pocket.
In that case, the carrier density will be less and come to a better agreement with the present estimates (0.7--0.9$\times10^{20}$ cm$^{-3}$) for $10.4 \leqslant P \leqslant 28.3$ kbar.

We now return to Fig.~\ref{color_plot}, which shows the pressure and temperature dependence of the carrier density $n_h$ and hole mobility $\mu_h$ based on analyses at all the measured $(P, T)$ points (indicated by crosses).
Changes associated with the structural transition at $T_s$ is moderate.
Although the estimated carrier density is gradually reduced below $T_s$, the temperature evolution of the mobility is hardly affected by the transition [Fig.~\ref{color_plot}(b)].
On the other hand, the AFM transition at $T_N$ is accompanied by a drastic reconstruction of the electronic structure, consistent with the high-pressure SdH study \cite{Terashima16PRB}.
The carrier density is reduced abruptly at $T_N$, suggesting that part of the FS is gapped.
The mobility is enhanced abruptly as well, which indicates that the carrier scattering due to spin fluctuations is suppressed.

It is intriguing that $T_c$ continues to increase with pressure despite the reduced carrier density and spin fluctuations in the AFM ordered phase.
Orbital fluctuations are likely reduced as well, since a structural distortion occurs at $T_N$ even when no separate structural transition exists at high pressures \cite{Kothapalli16condmat, Wang16condmat}.
Our analyses suggest the existence of a third tiny FS pocket both at low temperatures in the low-pressure paramagnetic phase and in the AFM phase.
The smallness of its Fermi energy, or the nearness of the band edge to the Fermi level, might deserve serious consideration in future studies \cite{Bianconi13NatPhys, Charnukha15SciRep, Terashima16PRB}. 

\begin{acknowledgments}
This work has been supported by a Grant-in-Aid for Scientific Research on Innovative Areas ``Topological Materials Science'' (KAKENHI Grant No. 15H05852) and JSPS KAKENHI Grant Number 26400373.
\end{acknowledgments}

%

\end{document}


\preprint{Ver. 3}¡¡

\title{Supplemental Material\\Magnetotransport study of the pressure-induced antiferromagnetic phase in FeSe}


\author{Taichi Terashima}
\author{Naoki Kikugawa}
\affiliation{National Institute for Materials Science, Tsukuba, Ibaraki 305-0003, Japan}
\author{Shigeru Kasahara}
\author{Tatsuya Watashige}
\author{Yuji Matsuda}
\affiliation{Department of Physics, Kyoto University, Kyoto 606-8502, Japan}
\author{Takasada Shibauchi}
\affiliation{Department of Advanced Materials Science, University of Tokyo, Chiba 277-8561, Japan}
\author{Shinya Uji}
\affiliation{National Institute for Materials Science, Tsukuba, Ibaraki 305-0003, Japan}


\date{\today}
\begin{abstract}
\end{abstract}


\maketitle



\newcommand{\ud}{\mathrm{d}}
\def\degree{\kern-.2em\r{}\kern-.3em}

\section{Sample characterization}
The chemical vapor transport (CVT) method can produce high-quality single crystals of FeSe.
B\"ohmer \textit{et al}. \cite{Bohmer13PRB} reported a composition of Fe$_{0.995(4)}$Se based on an analysis of X-ray diffraction.
Scanning tunneling microscopy by Watashige \textit{et al}. \cite{Watashige15PRX} found approximately only one defect per 5000 Fe atoms on cleaved surfaces of CVT-grown crystals.
So far we \cite{Terashima14PRB, Terashima16PRB, TerashimaUnpub} have performed SdH measurements on ten single crystals from different growth batches, and the standard deviations of the SdH frequencies are less than 1\%.
Since the frequencies are linked to the carrier density and hence chemical composition, this indicates that the composition of the grown crystals is highly reproducible.
Those crystals yield $T_c$ = 9.0(1) K, $T_s$ = 90.3(4) K, and the resistivity ratio at $T$ = 11 K of 30(3).
For comparison, the present sample gives 9.0 K, 90.5 K, and 29, respectively.
Thus the present sample is representative of CVT-grown high-quality single crystals, and hence measurements on it reveal intrinsic properties of high-quality FeSe.

\section{upper critical field}
\begin{figure}[b]
\includegraphics[width=8.6cm]{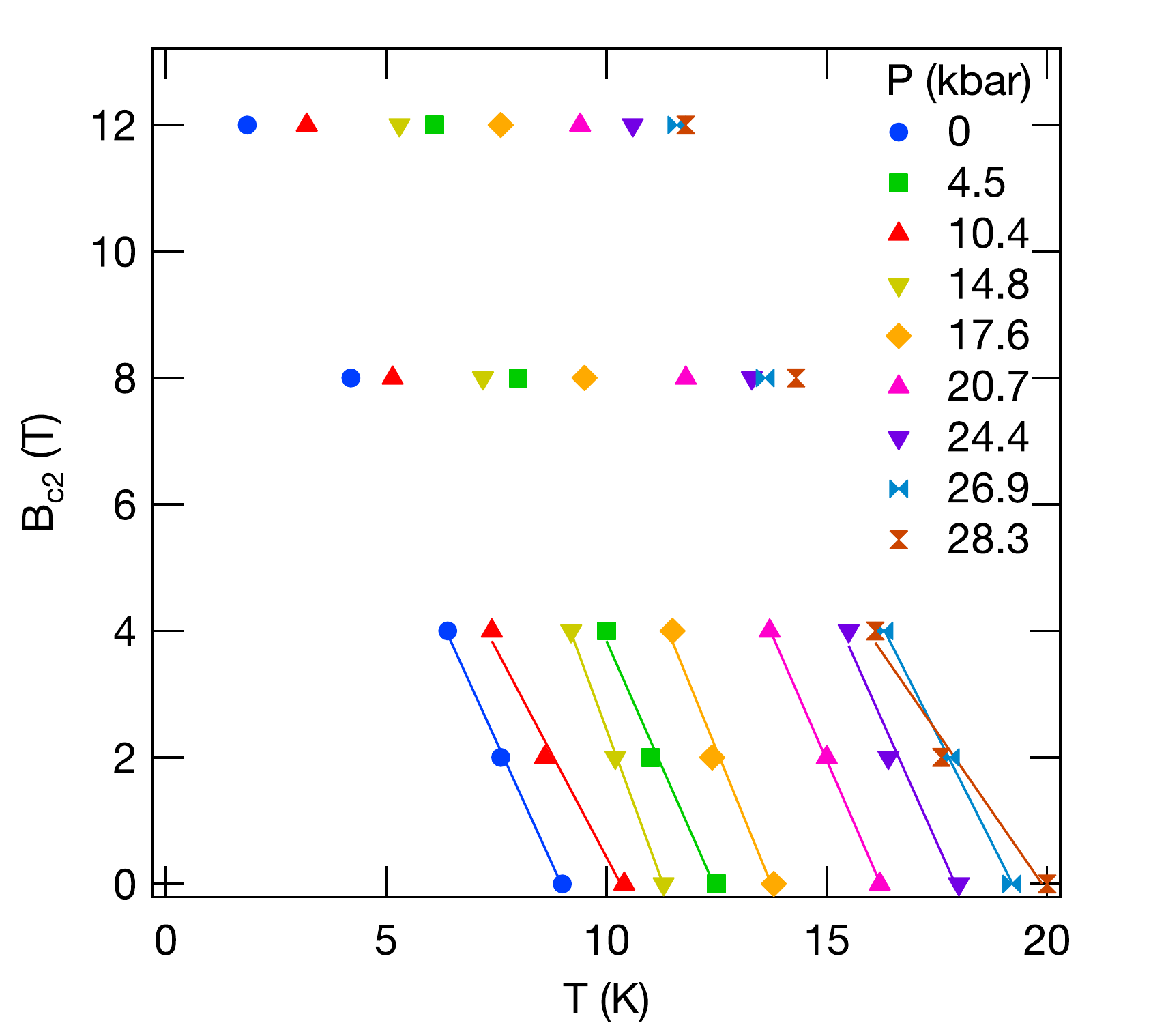}
\caption{\label{Bc2}(Color online).  Upper critical field $B_{c2}$ for $B \parallel c$ in FeSe as a function of temperature at various pressures.  $B_{c2}$ was determined by the zero-resistivity criterion.  Solid lines show linear fits to data up to $B$ = 4 T.}   
\end{figure}

\begin{figure}[b]
\includegraphics[width=8.6cm]{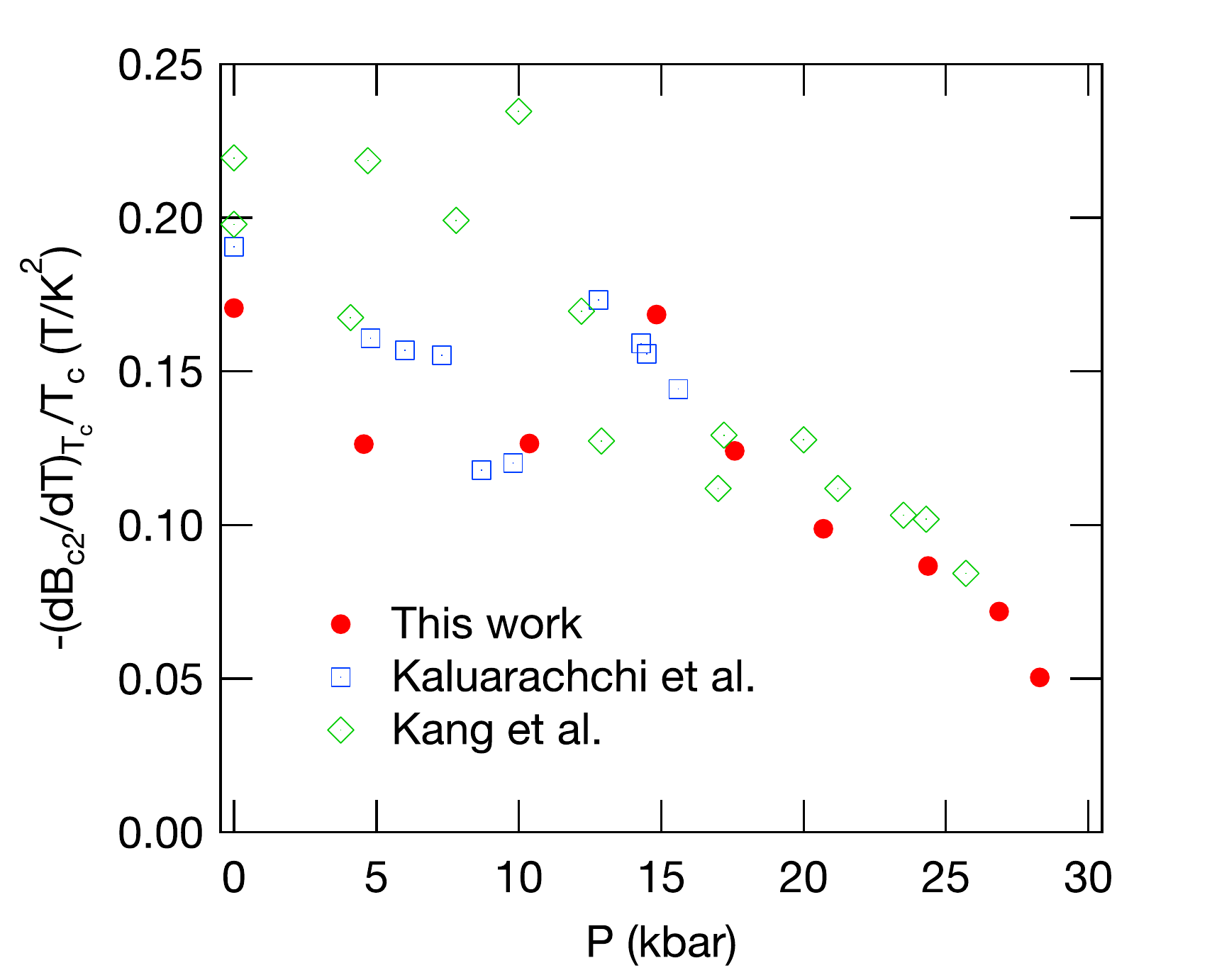}
\caption{\label{slope}(Color online).  Pressure dependence of the normalized initial slope $-$(d$B_{c2}$/d$T)_{T_c}$/$T_c$ for $B \parallel c$.  Data by Kaluarachchi \textit{et al}. \cite{Kaluarachchi16PRB} and Kang \textit{et al}. \cite{Kang16SST} are also shown.}   
\end{figure}

Figure~\ref{Bc2} shows the upper critical field $B_{c2}$ for $B \parallel c$ as a function of temperature at different pressures.
$B_{c2}$ was defined as the temperature where $\rho(T)$ at a constant field reaches zero resistivity.
Figure~\ref{slope} shows the normalized initial slopes $-$(d$B_{c2}$/d$T)_{T_c}$/$T_c$, which were determined from linear fits to data points up to $B$ = 4 T. 
The present and previous \cite{Kaluarachchi16PRB, Kang16SST}  data are in fair agreement.

\section{Multicarrier model}
We assume that the magnetic field $B$ is applied along the $z$ axis and that the current is in the $xy$ plane.
We consider the $xy$ components of the conductivity $\sigma_{\alpha\beta}$ and resistivity tensor $\rho_{\alpha\beta}$ ($\alpha$, $\beta$ = $x$, $y$).
We assign $\sigma^i_{\alpha\beta}$ to each $i$-th carrier:
\begin{equation}
\sigma^i_{\alpha\beta}= \frac{n_iq_i\mu_i}{1+(\mu_iB)^2}
\left( \begin{array}{cc}
1 & \mu_iB \\
-\mu_iB & 1 \\
\end{array} \right),
\end{equation}
where $n_i$, $q_i$, and $\mu_i$ are the density, charge, and mobility of the $i$-th carrier.
In our formalism, $q_i$, and $\mu_i$ have a sign (+ for holes and - electrons).
The conductivity and resistivity tensors are given by:
\begin{equation}
\sigma_{\alpha\beta}=\sum_i \sigma^i_{\alpha\beta},
\end{equation}
and
\begin{equation}
\rho_{\alpha\beta}=\sigma^{-1}_{\alpha\beta}.
\end{equation}
The relation to the experimental quantities $\rho$ and $\rho_H$ are as follow: $\rho$=$\rho_{xx}$, and $\rho_H$ = $\rho_{yx}$.

For the compensated two-carrier model, we have one hole ($i$ = h) and one electron ($i$ = e1) carrier and assume $n_h=n_{e1}$.
For the compensated three-carrier model, we have one hole ($i$ = h) and two electron ($i$ = e1 and e2) carriers and assume $n_h=n_{e1}+n_{e2}$.

%